\documentclass[10pt,twocolumn,letterpaper]{article}

\usepackage{cvpr}
\usepackage{times}
\usepackage{epsfig}
\usepackage{graphicx}
\usepackage{amsmath}
\usepackage{amssymb}
\graphicspath{ {./images/} }

\usepackage[pagebackref=true,breaklinks=true,letterpaper=true,colorlinks,bookmarks=false]{hyperref}

\cvprfinalcopy % *** Uncomment this line for the final submission

\ifcvprfinal\fi
\begin{document}

%%%%%%%%% TITLE
\title{Attention Based Image Compression Post-Processing Convolutional Neural Network}

\author{Yuyang Xue\\
University of Southampton\\
School of Electronics and Computer Science\\
{\tt\small yx2n17@soton.ac.uk}
\and
Jiannan Su\thanks{Corresponding author.}\\
Fuzhou University\\
Computer Science and Technology\\
{\tt\small sjn@fzu.edu.cn}
}

\maketitle
%\thispagestyle{empty}

%%%%%%%%% ABSTRACT
\begin{abstract}
   The traditional image compressors, e.g., BPG and H.266, have achieved great image and video compression quality. Recently, Convolutional Neural Network has been used widely in image compression. We proposed an attention-based convolutional neural network for low bit-rate compression to post-process the output of traditional image compression decoder. Across the experimental results on validation sets, the post-processing module trained by MAE and MS-SSIM losses yields the highest PSNR of 32.10 on average at the bit-rate of 0.15.
\end{abstract}

%%%%%%%%% BODY TEXT
\section{Introduction}

  Uncompressed image and video data require massive storage capacity and transmission bandwidth. For a long time, people are dreaming about a powerful compression method to greatly elevate the convenience of transmission and storage of applications and database. TBs or even PBs of data are consumed in the daily mobile network, most of which are videos and images. It is urgent to develop both practical and swift image and video compression technique to solve this problem. Typically, a traditional image compression method , like JPEG \cite{wallace1992jpeg} or JPEG2000\cite{christopoulos2000jpeg2000}, should go through DCT or wavelet transform, quantization, coding, entropy encoding, and decoding. While PNG \cite{boutell1997png}, and WebP \cite{ginesu2012objective} are widely used in daily life, people are still not very satisfied with the contemporary compression quality level. With the rise of deep learning, neural network has become a commonly used tool in various areas, mostly concerned with computer vision and natural language processing. Image compression, of course, has been successfully experimented that deep neural networks are effective in most situations. For example, autoencoder based neural network compression framework uses convolutional neural network stack to replace default feature extraction approaches on the traditional pipeline. GANs \cite{goodfellow2014generative}, are generally used to improve the MOS (mean opinion score). However, it will result in very low metric values in terms of PSNR and MS-SSIM if GANs are used.
  
  Our approach is to provide a fully convolutional neural network which is built upon attention mechanism to optimize the output of the traditional decoder, like BPG \cite{bellard} and H.266 \cite{jvet}. We found that the fixed threshold of the quality parameter can be further promoted by convolutional in both PSNR and MS-SSIM. One advantage of a fully convolutional neural network is that the input size of images can be arbitrary. The convolutional layer is also an expert in extracting features from the image.
  
 This paper is structured as follows: Section 2 covers related works in the area of image compression. The specification of our proposed method is discussed in Section 3. Section 4 shows the experimental results of the comparison between our methods and others. Finally, Section 5 concludes the characteristics of image compression techniques of previous sections as well as future work.
 \begin{figure*}
    \includegraphics[width=\textwidth]{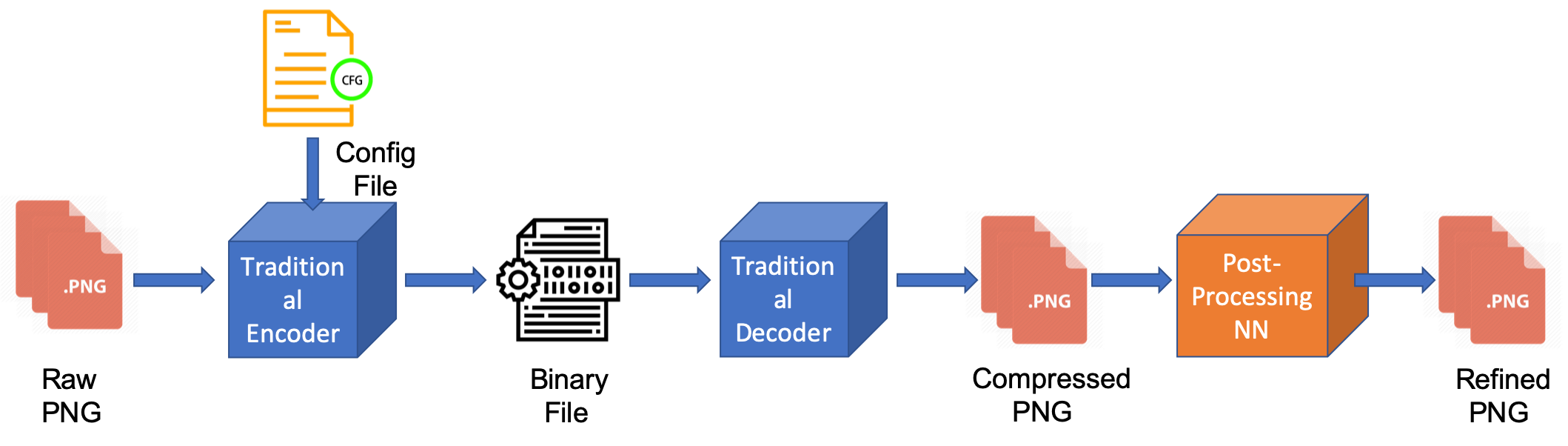}
    \caption{The proposed framework.}
    \label{fig:1}
\end{figure*}
\section{Related work}

As mentioned in the last chapter, image compression has gone through several generations. From traditional compression techniques like JPEG, JPEG2000, to recent adaptive image compression approach \cite{rippel2017real} proposed by Waveone, image compression evolves many research fields progress. Here are some related works:

\subsection{Traditional Image Compression Techniques}

JPEG is a widely used image format using 2D Fourier Discrete Cosine (DCT) Transform. It is also the foundation of most popular H.264 \cite{wiegand2003overview} video compression format. However, JPEG 2000 uses the wavelet transform to beat up JPEG for its higher quality in the same level of Bit Per Pixel (BPP). However, the lack of its application and slow encoding and decoding speed hinder its popularity. Google presented WebP in 2010 in order to substitute JPEG or PNG on the internet. The Predicting Module in the MacroBlocking of WebP sends the predicted output to the DCT transform, thus compressing the image in a more effective way. People tend to find a better way of image compression from video compression techniques. WebP can be seen as the key-frame compression of WebM \cite{bankoski2011intro}. BPG is derived from HEVC \cite{sullivan2012overview} (a.k.a the second part of H.265). It gives a higher dynamic range and a better compression ratio. H.266, i.e., Versatile Video Coding (VVC) \cite{institute} proposed by JVET group, is beyond H.265 and desire to have a preferable performance than any other traditional image/video compression methods.

\subsection{Deep Neural Networks}

Recent blossom of neural network in computer vision is highly attracted to researchers. Not only the convolutional layer but also recurrent structure may benefit image compression. GAN applications of image compression should be both perceptual and rational, but not be loyal to the source image.

\begin{itemize}
\item CNN and RNN:

In 2015, Toderici et al. \cite{toderici2015variable} proposed a creative architecture composed of convolutional and deconvolutional LSTM recurrent network. In order to get arbitrary size of input images, Toderici et al. \cite{toderici2017full} improved their framework into an RNN based end-to-end network. With the popularity and practicability of autoencoder, Theis et al. \cite{theis2017lossy} proposed compressive autoencoder with an upper-bound the discrete entropy rate loss for continuous relaxation. Jiang et al. \cite{jiang2018end} presented a fully convolutional end-to-end compression framework to work with residual and get the reconstructed image as similar to the groudtruth. Maleki et al. \cite{Aytekin_2018_CVPR_Workshops} proposed BlockCNN for artifact removal and achieved results with better quality on the enhanced images.

\item GAN: 

Only a few works are GAN based. Rippel et al. proposed a real-time adaptive image compression method \cite{rippel2017real} but the key is that target and reconstruction are no longer treated separately. They fine-tuned their GAN to decide when they propagated confusion signal, and when to train the discriminator. Galteri et al. \cite{galteri2017deep} presented a GAN based network with SSIM loss to get a better artifact removal result. Lomnitz et al. \cite{squeezedin} proposed Compression-GAN to re-generate human faces in rather low szies while with high MOS score.

\end{itemize}

\subsection{Traditional and DNN Fused}

Traditional compression techniques have their advantage of fast encoding/decoding speed, and without plenty of data to train. Liu et al. \cite{Liu_2018_CVPR_Workshops} combined JPEG with a deep neural network to ease the storage and data communication overhead. Chen et al. \cite{Chen_2018_CVPR_Workshops} presented a CNN-optimized image compression with uncertainty based resource allocation. They used a CNN based method to predict the probability distribution of syntax element and boost the performance of in-loop filtering with a novel convolutional network that incorporates dense connections and identity skip connections. Their team gained the first prize of the 2018 CVPR compression workshop.

%-------------------------------------------------------------------------

\section{Method}

In this study, we develop our codec based on the Versatile Video Coding (VVC) \cite{institute}, which is the codec developed based on H.266 structure. In addition, we designed a post-processing method, which is based on convolutional neural network to improve coding performance. An overview of the image compression framework is depicted in Fig. 1. The proposed framework mainly consists of two parts: traditional codec and post-processing. The network structure used in the post-processing stage is shown in Fig. 2. The network is composed of 30 residual blocks, each of which contains channel attention (CA) and spatial attention (SA)  mechanisms \cite{Chen_2017_CVPR}.

Given a set of compressed images $\{\mathbf{X_i}\}$ and their corresponding ground truth images $\{\mathbf{Y_i}\}$, we use mean absolute error (MAE) as the loss function:
\begin{equation}
L(\Theta) = \frac{1}{N}\sum_{i=1}^{N}|f(\mathbf{X_i};\Theta)-\mathbf{Y_i}|
\end{equation}
where $\Theta$ is the network parameters. $f$ is the mapping function. $N$ is the number of training samples.

After 10 thousand iterations, we used a combination of both MAE and MS-SSIM as our losses to elevate MS-SSIM score. The MS-SSIM can be presented as:

\begin{equation}
L_{MS-SSIM} = [L_M(X, Y)]^{\alpha_M}\prod ^{M}_{J=1}[C_J(X, Y)]^{\beta_j}[S_J(X, Y)]^{\gamma_J}
\end{equation}

where $X, Y$ are the pixels of the image. The SSIM formula is based on three comparison measurements between the samples of $X$ and $Y$: luminance $L$, contrast $C$ and structure $S$. The MS-SSIM is changing the image scale with the coefficient M. Therefore, the whole loss function is:

\begin{equation}
L = \lambda L_{MS-SSIM} + L(\Theta)
\end{equation}

where $\lambda$ is set to $\frac{1}{20}$ in our experiment.

 \begin{figure}
    \includegraphics[width=\linewidth]{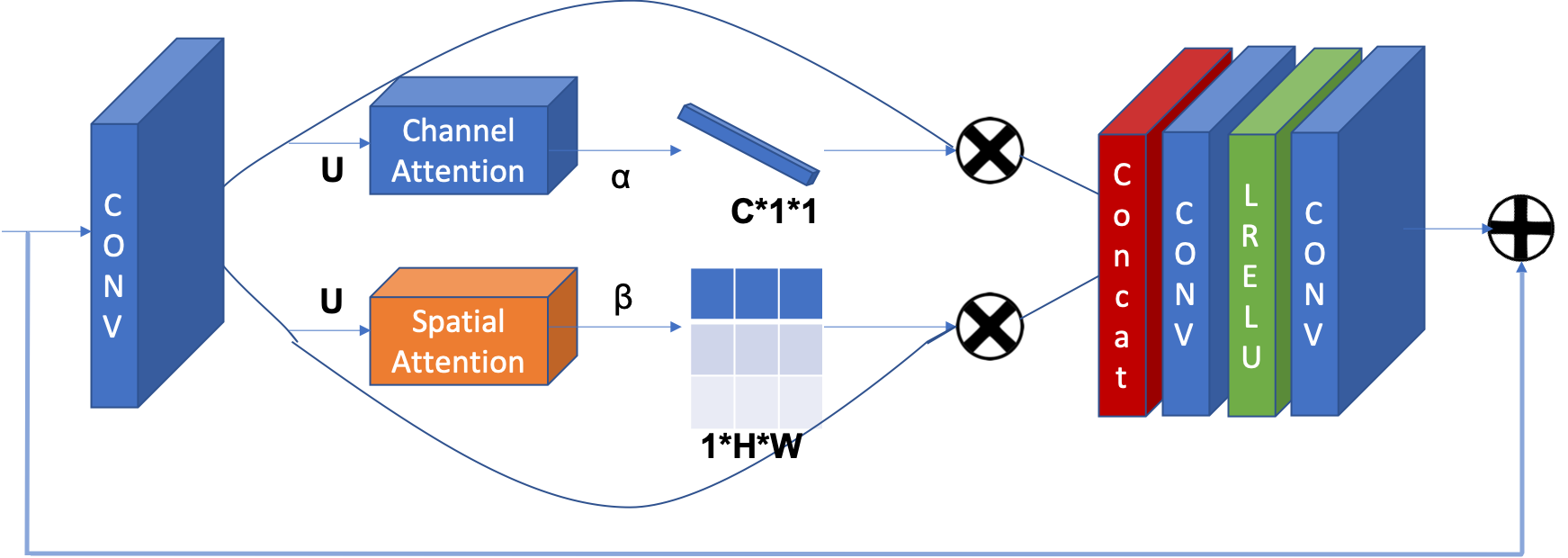}
    \caption{Attention Residual Block.}
    \label{fig:2}
\end{figure}
\section{Experiment}

  This section describes datasets that we used, the main training procedure and hyper-parameter settings.
  
\subsection{Datasets}

Since we have evaluated the size and distribution of the provided training set, we choose DIV2K dataset \cite{Ignatov_2018_ECCV_Workshops} as our main training data. Moreover, the provided training set contains two part of data, mobile and professional. To fit better the mobile training set, we choose DPED dataset \cite{ignatov2017dslr}. We equally mixed two datasets together as our main training data. For training, we randomly crop $64\times 64$, $128\times 128$ and $256\times 256$ patches from $90\%$ of these images as inputs to the traditional encoder in a fixed quality parameter. We only use rotation as augmentation. Then, after decoding all the output of the encoder, we sent the decoded result to the network for training. For testing and validation, we used the other $10\%$.

\subsection{Procedure}

It is indispensable to test several traditional compressors as we aimed to develop a post-processing network. BPG and H.266 image compressor stand out in terms of PSNR and MS-SSIM metrics, so we choose them as the baseline of the test. The experiment was carried out in 3 phases. For the first phase, we searched a range of quality parameters for the encoder to compress images to an average of 0.15 bpp. The bpp is different due to the fact that the fixed quality parameter can not always get the exact size of image. We have to mix the adjacent quality parameter output to get the closest result. After that, the decoder gets the outputs of binary data as neural networks inputs. At the third phase, we tested the different crop sizes of inputs to choose the best fit. We adapted EDSR+'s \cite{lim2017enhanced} method to rotate the image in every 90 degrees, and we put every rotated image to the model to get the average of inferenced outputs to improve the PSNR and MS-SSIM indices.

\subsection{Implementation details}
The learning rates for the network was set to  $1e^{-4}$.  We use the Adam and set $\beta_{1} = 0$, $\beta_{2}= 0.999$. All the experiments are conducted on a P5000 GPU with Intel 7600 CPU. We have set the limit to 12GB RAM according to the rule of the competition. We set the $\lambda$ of the loss function to $0.05$. The total decoding time on this machine is one and half an hour on the whole validation set. The rotation version of the post-process takes about six hours.

\subsection{Results}

In order to get the max of 0.15 bpp limitation, we have to mix results from different quality parameters. We choose BPG and H.266 as baselines. For BPG, we mixed quality parameters at 40 and 41; As for H.266, we mixed quality parameters at 35, 36 and 37 for different configure setting. As shown in Table 1, it can be seen the post-processing module improves the PSNR and MS-SSIM metrics for both algorithms. The EDSR's rotation trick elevates about 1 percent of PSNR and 0.2 percent of MS-SSIM. We have submitted as the name of $ColorBlust$ as our final result.

\begin{table}[]
\begin{tabular}{|l|c|c|c|}
\hline
Approachs               & PSNR  & MS-SSIM & BPP \\ \hline
BPG                     & 31.47 & 0.94824 &  0.144   \\ \hline
BPG+Post                & 32.01 & 0.95712 &  0.148   \\ \hline
H.266                   & 31.72 & 0.96097 &  0.149   \\ \hline
H.266 + Post            & 32.09 & 0.96104 &  0.147   \\ \hline
H.266 + Post + Rotation & 32.10 & 0.96124 &  0.149   \\ \hline
\end{tabular}
\caption { Evaluation results on CLIC 2019 validation dataset}
\end{table}

\section{Conclusion}

In this work, we have presented a post-processing fully convolutional neural network for the decoder to improve its objective evaluation metrics. The post-processing performance is mainly depended on the quality of the front-end traditional compressor. A post-processing network has targeted to optimize PSNR or MS-SSIM. The convolutional neural network maybe slow for elevating image quality but it is worthwhile for promoting PSNR for 0.64 dB. In fact, the more reliable the front-end compressor is, the higher quality we can get. In future work, it would be interested to explore a more effective network and make a convolutional neural network into traditional encoder to optimize its performance.

{\small
\bibliographystyle{ieee}
\bibliography{egbib}

\begin{thebibliography}{10}\itemsep=-1pt

\bibitem{jvet}
Developing a video compression algorithm with capabilities beyond hevc.

\bibitem{Aytekin_2018_CVPR_Workshops}
C.~Aytekin, X.~Ni, F.~Cricri, J.~Lainema, E.~Aksu, and M.~Hannuksela.
\newblock Block-optimized variable bit rate neural image compression.
\newblock In {\em The IEEE Conference on Computer Vision and Pattern
  Recognition (CVPR) Workshops}, June 2018.

\bibitem{bankoski2011intro}
J.~Bankoski.
\newblock Intro to webm.
\newblock In {\em Proceedings of the 21st international workshop on Network and
  operating systems support for digital audio and video}, pages 1--2. ACM,
  2011.

\bibitem{bellard}
F.~Bellard.
\newblock Better portable graphics.

\bibitem{boutell1997png}
T.~Boutell.
\newblock Png (portable network graphics) specification version 1.0.
\newblock Technical report, 1997.

\bibitem{Chen_2017_CVPR}
L.~Chen, H.~Zhang, J.~Xiao, L.~Nie, J.~Shao, W.~Liu, and T.-S. Chua.
\newblock Sca-cnn: Spatial and channel-wise attention in convolutional networks
  for image captioning.
\newblock In {\em The IEEE Conference on Computer Vision and Pattern
  Recognition (CVPR)}, July 2017.

\bibitem{Chen_2018_CVPR_Workshops}
Z.~Chen, Y.~Li, F.~Liu, Z.~Liu, X.~Pan, W.~Sun, Y.~Wang, Y.~Zhou, H.~Zhu, and
  S.~Liu.
\newblock Cnn-optimized image compression with uncertainty based resource
  allocation.
\newblock In {\em The IEEE Conference on Computer Vision and Pattern
  Recognition (CVPR) Workshops}, June 2018.

\bibitem{christopoulos2000jpeg2000}
C.~Christopoulos, A.~Skodras, and T.~Ebrahimi.
\newblock The jpeg2000 still image coding system: an overview.
\newblock {\em IEEE transactions on consumer electronics}, 46(4):1103--1127,
  2000.

\bibitem{galteri2017deep}
L.~Galteri, L.~Seidenari, M.~Bertini, and A.~Del~Bimbo.
\newblock Deep generative adversarial compression artifact removal.
\newblock In {\em Proceedings of the IEEE International Conference on Computer
  Vision}, pages 4826--4835, 2017.

\bibitem{ginesu2012objective}
G.~Ginesu, M.~Pintus, and D.~D. Giusto.
\newblock Objective assessment of the webp image coding algorithm.
\newblock {\em Signal Processing: Image Communication}, 27(8):867--874, 2012.

\bibitem{goodfellow2014generative}
I.~Goodfellow, J.~Pouget-Abadie, M.~Mirza, B.~Xu, D.~Warde-Farley, S.~Ozair,
  A.~Courville, and Y.~Bengio.
\newblock Generative adversarial nets.
\newblock In {\em Advances in neural information processing systems}, pages
  2672--2680, 2014.

\bibitem{ignatov2017dslr}
A.~Ignatov, N.~Kobyshev, R.~Timofte, K.~Vanhoey, and L.~Van~Gool.
\newblock Dslr-quality photos on mobile devices with deep convolutional
  networks.
\newblock In {\em Proceedings of the IEEE International Conference on Computer
  Vision}, pages 3277--3285, 2017.

\bibitem{Ignatov_2018_ECCV_Workshops}
A.~Ignatov, R.~Timofte, et~al.
\newblock Pirm challenge on perceptual image enhancement on smartphones:
  report.
\newblock In {\em European Conference on Computer Vision (ECCV) Workshops},
  January 2019.

\bibitem{institute}
F.~H.~H. Institute.
\newblock Versatile video coding (vvc).

\bibitem{jiang2018end}
F.~Jiang, W.~Tao, S.~Liu, J.~Ren, X.~Guo, and D.~Zhao.
\newblock An end-to-end compression framework based on convolutional neural
  networks.
\newblock {\em IEEE Transactions on Circuits and Systems for Video Technology},
  28(10):3007--3018, 2018.

\bibitem{lim2017enhanced}
B.~Lim, S.~Son, H.~Kim, S.~Nah, and K.~Mu~Lee.
\newblock Enhanced deep residual networks for single image super-resolution.
\newblock In {\em Proceedings of the IEEE Conference on Computer Vision and
  Pattern Recognition Workshops}, pages 136--144, 2017.

\bibitem{Liu_2018_CVPR_Workshops}
H.~Liu, T.~Chen, Q.~Shen, T.~Yue, and Z.~Ma.
\newblock Deep image compression via end-to-end learning.
\newblock In {\em The IEEE Conference on Computer Vision and Pattern
  Recognition (CVPR) Workshops}, June 2018.

\bibitem{squeezedin}
M.~Lomnitz.
\newblock Squeezed-in extreme image compression with deep learning.
\newblock Technical report, 2017.

\bibitem{rippel2017real}
O.~Rippel and L.~Bourdev.
\newblock Real-time adaptive image compression.
\newblock In {\em Proceedings of the 34th International Conference on Machine
  Learning-Volume 70}, pages 2922--2930. JMLR. org, 2017.

\bibitem{sullivan2012overview}
G.~J. Sullivan, J.-R. Ohm, W.-J. Han, and T.~Wiegand.
\newblock Overview of the high efficiency video coding (hevc) standard.
\newblock {\em IEEE Transactions on circuits and systems for video technology},
  22(12):1649--1668, 2012.

\bibitem{theis2017lossy}
L.~Theis, W.~Shi, A.~Cunningham, and F.~Husz{\'a}r.
\newblock Lossy image compression with compressive autoencoders.
\newblock {\em arXiv preprint arXiv:1703.00395}, 2017.

\bibitem{toderici2015variable}
G.~Toderici, S.~M. O'Malley, S.~J. Hwang, D.~Vincent, D.~Minnen, S.~Baluja,
  M.~Covell, and R.~Sukthankar.
\newblock Variable rate image compression with recurrent neural networks.
\newblock {\em arXiv preprint arXiv:1511.06085}, 2015.

\bibitem{toderici2017full}
G.~Toderici, D.~Vincent, N.~Johnston, S.~Jin~Hwang, D.~Minnen, J.~Shor, and
  M.~Covell.
\newblock Full resolution image compression with recurrent neural networks.
\newblock In {\em Proceedings of the IEEE Conference on Computer Vision and
  Pattern Recognition}, pages 5306--5314, 2017.

\bibitem{wallace1992jpeg}
G.~K. Wallace.
\newblock The jpeg still picture compression standard.
\newblock {\em IEEE transactions on consumer electronics}, 38(1):xviii--xxxiv,
  1992.

\bibitem{wiegand2003overview}
T.~Wiegand, G.~J. Sullivan, G.~Bjontegaard, and A.~Luthra.
\newblock Overview of the h. 264/avc video coding standard.
\newblock {\em IEEE Transactions on circuits and systems for video technology},
  13(7):560--576, 2003.

\end{thebibliography}
}

\end{document}